\documentclass[a4paper,fleqn,usenatbib,useAMS]{mnras}
\usepackage{graphicx}   
\usepackage{color}
\usepackage{lscape}
\usepackage{txfonts}
\usepackage{amsfonts}

\title[Radio/X-ray correlation in three NS-LMXBs]
{A model for the radio/X-ray correlation in three neutron star low-mass X-ray binaries 4U 1728-34, 
Aql X-1 and EXO 1745-248}
\author[Erlin Qiao and B.F. Liu]{Erlin Qiao $^{1,2}$\thanks{E-mail:
qiaoel@nao.cas.cn} and B.F. Liu $^{1,2}$\\
$^{1}$Key Laboratory of Space Astronomy and Technology, National Astronomical Observatories, Chinese Academy of
Sciences, Beijing 100012, China \\
$^{2}$School of Astronomy and
Space Sciences, University of Chinese Academy of Sciences, 19A Yuquan Road, Beijing 100049, China\\} 

\date{Accepted XXX. Received YYY; in original form ZZZ}
\pubyear{2019}

\begin{document}
\label{firstpage}
\pagerange{\pageref{firstpage}--\pageref{lastpage}}
\maketitle
\begin{abstract}
Observationally, for neutron star low-mass X-ray binaries,  
so far, the correlation between the radio luminosity $L_{\rm R}$  and the X-ray luminosity 
$L_{\rm X}$, i.e., $L_{\rm R}\propto L_{\rm X}^{\beta}$, has been  reasonably well-established only 
in three sources 4U 1728-34, Aql X-1 and  EXO 1745-248 in their hard state. 
The slope $\beta$ of the radio/X-ray correlation of the three 
sources is different, i.e., $\beta \sim 1.4$ for 4U 1728-34, $\beta \sim 0.4$ for Aql X-1, 
and $\beta \sim 1.6$ for EXO 1745-248. In this paper, for the first time we explain the different 
radio/X-ray correlation of 4U 1728-34, Aql X-1 and  EXO 1745-248 with the coupled 
advection-dominated accretion (ADAF)-jet model respectively. 
We calculate the emergent spectrum of the ADAF-jet model for $L_{\rm X}$ and $L_{\rm R}$ at 
different $\dot m$ ($\dot m=\dot M/\dot M_{\rm Edd}$), adjusting $\eta$ 
($\eta \equiv \dot M_{\rm jet}/\dot M$, describing the fraction of the accreted matter 
in the ADAF transfered vertically forming the jet) to fit the observed radio/X-ray correlations. 
Then we derive a fitting formula of $\eta$ as a function of $\dot m$ for 4U 1728-34, Aql X-1 and 
EXO 1745-248 respectively. If the relation between $\eta$ and $\dot m$ can be 
extrapolated down to a lower value of $\dot m$, we find that in a wide range of $\dot m$,  
the value of $\eta$ in Aql X-1 is greater than that of in 4U 1728-34 and EXO 1745-248, 
implying that Aql X-1 may have a relatively stronger large-scale magnetic field, 
which is supported by the discovery of the coherent millisecond  X-ray pulsation 
in Aql X-1. 
\end{abstract}




\section{Introduction}
Low-mass X-ray binaries (LMXBs) which either contain a black hole (BH) or a neutron star (NS), 
accreting matter from its low-mass companion star ($\lesssim 1 M_{\odot}$) are  ideal  
natural laboratories for studying the physics of accretion and jet around a BH or a NS 
\citep[e.g.][]{Migliari2006}. According to the timing and spectral features in the X-ray band,
LMXBs are generally divided into two main spectral states, i.e., the high/soft state
and the low/hard state \citep[][for review]{Gilfanov2010}.
For BH-LMXBs, when they are in the high/soft state, the accretion flow is 
widely believed to be dominated by the optically thick, geometrically thin, cool 
accretion disc \citep[][]{Shakura1973}, and the X-ray spectrum can be well described 
by a multi-color blackbody spectrum \citep[e.g.][]{Mitsuda1984,Makishima1986,Merloni2000}.
Whereas, for NS-LMXBs, besides the emission from the disc, there is a significant thermal  
emission from the boundary layer between the accretion disc and the surface of the 
NS \citep[][for review]{Popham1992,Inogamov1999,Popham2001,Gilfanov2014}. 
Observationally, for both BH-LMXBs and NS-LMXBs, when they are in the low/hard state, generally, 
the accretion flow is suggested to be dominated by the optically thin, 
geometrically thick, hot, advection-dominated accretion flow (ADAF) \citep[][for review]{Done2007}.
Theoretically, the ADAF solution has been studied in detail by several researchers 
since it was discovered in 1970's \citep[][for review]{Ichimaru1977,Rees1982,Narayan1994,
Narayan1995a,Narayan1995b,Abramowicz1995,Chen1995,Yuan2014}. 
In the BH case, the ADAF solution is a kind of radiatively inefficient accretion flow, 
in which a fraction of the viscously dissipated energy will be advected into the event horizon of the BH. 
While in the NS case, the viscously dissipated energy advected onto the surface of the NS will 
eventually be radiated out, so the ADAF solution is radiatively efficient \citep[][]{Narayan1995b}. 
\citet[][]{Qiao2018b} calculated the structure and the corresponding emergent spectrum of 
the ADAF around a weakly magnetized NS within the framework of the  self-similar solution of 
the ADAF. The authors compared the electron temperature of the ADAF around a NS and a BH, 
it is found that the electron temperature of the ADAF around a NS is systemically lower 
than that of a BH, which is consistent with observations \citep[][]{Burke2017,Qiao2018b}.
Meanwhile, the authors compared the Compton $y$-parameter 
(defined as $y={{4kT_{\rm e}}\over {m_{\rm e}c^2}} \rm {Max}(\tau_{\rm es}, \tau^2_{\rm es})$,
with $T_{\rm e}$ being the election temperature, $m_{\rm e}$ being the electron mass,
$c$ being the speed of light, and $\tau_{\rm es}$ being the Compton scattering optical depth)
of the ADAF around a NS and a BH, it is found that the Compton $y$-parameter of the ADAF 
around a NS is systemically lower than that of a BH, producing a softer X-ray 
spectrum, which is also consistent with observations 
\citep[][]{Wijnands2015,Parikh2017,Sonbas2018,Qiao2018b}.

Observationally, an empirical correlation between the radio luminosity and the 
X-ray luminosity has been established in the low/hard state of several BH-LMXBs 
with a form of $L_{\rm R}\propto L_{\rm X}^{\sim 0.5-0.7}$ in the range of the X-ray 
luminosity from $\sim 10^{32}$ to $\sim 10^{38}\ \rm erg \ s^{-1}$, 
which is generally called the `universal' correlation 
\citep[e.g.][]{Hannikainen1998,Corbel2000,Corbel2003,Gallo2003,Corbel2008,Corbel2013}.
We should be careful that a significant deviation from the 
`universal' radio/X-ray correlation is observed for some sources 
(such as GX 339-4) in some single outburst
or decay in the X-ray luminosity range of $\sim 1$ order of magnitude 
\citep[][]{Corbel2013}.
\citet[][]{Yuan2005b} interpreted the radio/X-ray correlation of 
$L_{\rm R}\propto L_{\rm X}^{\sim 0.5-0.7}$ with the radiatively inefficient 
accretion flow, i.e.,  ADAF, + jet model, in which the X-ray emission is dominated 
by the ADAF and the radio emission is dominated by the jet. 
Recently, a growing number of sources have been discovered to tend to cluster
below the `universal' correlation with a steeper correlation of 
$L_{\rm R}\propto L_{\rm X}^{0.98\pm 0.08}$, which is possibly indicative of 
two distinct tracks \citep[][]{Gallo2012}.  
The study for BH-LMXB H1743-322 shows that there is a correlation of 
$L_{\rm R}\propto L_{\rm X}^{\sim 1.4}$ in the X-ray luminosity range from 
$\sim 10^{36}-10^{38}\ \rm erg \ s^{-1}$, and H1743-322 transits to the  
`universal' correlation as it fades towards the quiescence \citep[][]{Coriat2011}.
It is suggested that the correlation of $L_{\rm R}\propto L_{\rm X}^{\sim 1.4}$ may be 
resulted by the transition of the accretion flow from the radiatively inefficient accretion 
flow, ADAF, to the radiatively efficient accretion flow, disc-corona system, with increasing the 
mass accretion rate \citep[][for discussions]{Coriat2011,Gallo2012}, which is later 
quantitatively interpreted by several authors with the disc-corona + jet model 
\citep[e.g.][]{Qiao2015,Huang2014,MeyerHofmeister2014}. 
However, we note that the statistical significance of the aforementioned 
two tracks, i.e., the  `universal' correlation and the steeper 
correlation is still questioned \citep[][]{Gallo2014,Gallo2018}. 

For NS-LMXBs, so far, individually, the radio/X-ray correlation has been  reasonably 
well-established only in three sources in the X-ray luminosity range of 
$\sim 10^{36}-10^{37} \rm \ erg \ s^{-1}$ , i.e., 4U 1728-34 \citep[][]{Migliari2003}, 
Aql X-1 \citep[][]{Tudose2009,Miller-Jones2010}, and  EXO 1745-248 \citep[][]{Tetarenko2016}. 
By fitting the data (including both the high/soft state and the low/hard state ) of 4U 1728-34 with 
12 simultaneous Very Large Array (VLA) and {\it Rossi X-ray Timing Explorer} (RXTE) observations 
between 2000 and 2001, it is found that the 
radio/X-ray correlation of 4U 1728-34 is $L_{\rm R}\propto L_{\rm X}^{\sim 1.4}$,
which is consistent with what is expected by the radiatively efficient accretion flow due 
to the existence of the surface of the NS \citep[][]{Migliari2003}. 
The fitting result for the radio/X-ray correlation of Aql X-1 is 
$L_{\rm R}\propto L_{\rm X}^{\sim 0.4}$, which is closer
to the prediction by the radiatively inefficient accretion flow \citep[][]{Tudose2009}.
However, we should keep in mind that the data used in \citet[][]{Tudose2009}
also include the data of both the high/soft state and the low/hard state as \citet[][]{Migliari2003}.
\citet[][]{Migliari2006} jointly fitted the data of 4U 1728-34 and Aql X-1 only in 
the low/hard state, the authors found that the radio/X-ray correlation is 
$L_{\rm R}\propto L_{\rm X}^{\sim 1.4}$ (only two points for Aql X-1 are included).
\citet[][]{Tetarenko2016} analyzed the near-simultaneous data of VLA,
Australia Telescope Compact Array (ATCA), and $\it Swift$ X-ray Telescope in the low/hard
state of EXO 1745-248, located in the globular cluster Terzan 5, it is 
found that the radio/X-ray correlation of EXO 1745-248 is 
$L_{\rm R}\propto L_{\rm X}^{\sim 1.68}$.

In this paper, we focus on the radio/X-ray correlation in the low/hard 
state of the three NS-LMXBs 4U 1728-34, Aql X-1 and  EXO 1745-248. 
Specifically, we collect the simultaneous radio (8.5 GHz)  
and X-ray (2-10 keV) data of 4U 1728-34 \citep[][]{Migliari2003}, Aql X-1 
\citep[][]{Migliari2006,Tudose2009,Miller-Jones2010}, and EXO 1745-248 \citep[][]{Tetarenko2016}
in the low/hard state from literatures. We show that there is a correlation of 
$L_{\rm R}\propto L_{\rm X}^{1.4}$ for 4U 1728-34, 
a correlation of $L_{\rm R}\propto L_{\rm X}^{0.4}$ for Aql X-1, and 
a correlation of $L_{\rm R}\propto L_{\rm X}^{1.6}$ for 
EXO 1745-248 (note: the slope $\beta$ here is 1.6, which is little different from 
the value of 1.68 in \citet[][]{Tetarenko2016}. Please refer to Section \ref{sec:EXO 1745-248}
for the reason). Then we modelled the radio/X-ray correlation of 4U 1728-34, Aql X-1 
and EXO 1745-248 within the framework of the coupled ADAF-jet model respectively.  
We follow \citet[][]{Qiao2018b} for calculating the structure and the emergent spectrum of the ADAF around 
a weakly magnetized NS. In the coupled ADAF-jet model, we define a parameter, $\eta$, connecting the ADAF
and the jet. Specifically, $\eta \equiv \dot M_{\rm jet}/\dot M$, 
(with $\dot M$ being the mass accretion rate in the ADAF and $\dot M_{\rm jet}$ being 
the mass rate in the jet), is defined to describe 
the fraction of the mass accretion rate in the ADAF transfered vertically forming the jet. 
We calculate the emission of the jet as assumed in the internal shock scenario \citep[e.g.][]{Yuan2005}. 
We calculate the emergent spectrum of the coupled ADAF-jet model for $L_{\rm X}$ and $L_{\rm R}$ at different 
$\dot m$ (with $\dot m=\dot M/\dot M_{\rm Edd}$,  
and $\dot M_{\rm Edd}$ = $1.39 \times 10^{18} M/M_{\rm \odot} \rm \ g \ s^{-1}$),
adjusting the value of $\eta$ to fit the radio/X-ray correlation of 4U 1728-34, Aql X-1 
and EXO 1745-248 respectively. Then we derive a fitting formula between $\eta$ and $\dot m$ 
for 4U 1728-34, Aql X-1 and EXO 1745-248 respectively. 
Such a relation between $\eta$ and $\dot m$ may provide some clues on the formation mechanism and power of 
the jet in NS-LMXBs. Finally, we extrapolate the relation between  $\eta$ and $\dot m$ down to 
a lower value of  $\dot m$, predicting a new value of X-ray luminosity and radio luminosity, 
which we expect could be confirmed by the observations in the future.
The paper is organised as follows. The coupled ADAF-jet model is briefly introduced and discussed 
in Section 2. The results are shown in Section 3. The discussions are in Section 4 
and conclusions are in Section 5.

\section{The model}
\subsection{The ADAF around a weakly magnetized NS}\label{ADAF}
We follow the paper of \citet[][]{Qiao2018b} for the structure and the emergent spectrum of 
the ADAF around a weakly magnetized NS. The self-similar solution of the ADAF is adopted
in \citet[][]{Qiao2018b} as first proposed by \citet[][]{Narayan1995b}.
Specifically, we consider that the internal energy stored in the ADAF and the radial 
kinetic energy of the ADAF are transfered onto the surface of the NS.
It is assumed that a fraction, $f_{\rm th}$, of this energy is thermalized at 
the surface of the NS as the soft photons to be scattered in the ADAF. We self-consistently 
calculate the structure of the emergent spectrum of the ADAF by considering the 
radiative coupling between the soft photons from the surface of the NS and ADAF itself.
The structure and the corresponding emergent spectrum of the ADAF can be calculated
by specifying the mass of the NS $m$ ($m=M/M_{\odot}$), 
the mass accretion rate $\dot m$, the radius of the NS $R_{*}$ (i.e., the inner 
boundary of the ADAF), the viscosity parameter $\alpha$, the magnetic parameter 
$\beta^{'}$ (with magnetic pressure $p_{\rm m}={B^2/{8\pi}}=(1-\beta^{'})p_{\rm tot}$,
$p_{\rm tot}=p_{\rm gas}+p_{\rm m}$), and $f_{\rm th}$. As we can see the analyses in  
\citet[][]{Qiao2018b} and \citet[][]{Narayan1995b}, there is a critical mass accretion 
rate $\dot M_{\rm crit}$, above which the ADAF solution cannot exist. For accreting NS,
the critical mass accretion rate is $\dot M_{\rm crit} \sim 0.1\alpha^2\dot M_{\rm Edd}$.
Assuming a typical NS mass of $m=1.4$, the corresponding critical luminosity 
is $L_{\rm crit} \sim 0.1\alpha^2 L_{\rm Edd}\sim 0.1\alpha^2
\times 10^{38} \rm \ erg \ s^{-1}$ (with $L_{\rm Edd}=1.26\times 10^{38}m \rm \ erg \ s^{-1}$ 
and assuming that the radiative efficiency of the ADAF is 0.1).
In order to match the observed upper limits of the X-ray luminosity in the hard state (approaching or exceeding 
$10^{37} \rm \ erg \ s^{-1}$) of 4U 1728-34, Aql X-1 and EXO 1745-248 for the 
radio/X-ray correlation, we have to  choose a bigger value of $\alpha$. Throughout the paper,
we set $\alpha=1$ as suggest by \citet[][]{Narayan1996} for the luminous hard state of 
BH X-ray binaries. The ADAF solution often has a relatively weak magnetic field, as suggested 
by magnetohydrodynamic simulations \citep[][for review]{Yuan2014}. Throughout the paper,
we set $\beta^{'}=0.95$ \citep[][]{Qiaoetal2013,Qiao2018a}. Meanwhile, it is unclear how the energy 
of the accretion flow transfered onto the surface of the NS will interact with the NS, we simply assume 
$f_{\rm th}=1$ throughout the paper, meaning that all the energy advected onto the 
surface of the NS is thermalized as the soft photons to be scattered in the ADAF.  
So, we have three free parameters left,
i.e., the NS mass $m$, the NS radius $R_{*}$, and the mass accretion rate $\dot m$.

\subsection{The coupled ADAF-jet model}
The emission of the jet is calculated as assumed in the internal shock scenario, which has been widely
used to interpret the radio emission in radio-loud quasars \citep[][]{Spada2001}, 
in low-luminosity active galactic nuclei \citep[e.g.][]{Yu2011,Nemmen2014,vanOers2017}, 
in the Galactic center Sgr A* \citep[][]{Yuan2002,Ressler2017}, in the low/hard 
state of BH-LMXBs \citep[][]{Yuan2005,Zhang2010,Xie2016}, and in gamma-ray burst (GRB)  
afterglow \citep[e.g.][]{Piran1999} etc.
In the internal shock scenario, the jet emission can be calculated by specifying the parameters
as follows. (1) the mass rate $\dot M_{\rm jet}$ in the jet.
(2) the half-opening angle of the jet $\phi$ (assuming the jet with a conical geometry). 
(3) the bulk Lorentz factor of the jet $\Gamma_{\rm jet}$. (4) $\epsilon_{\rm e}$ and $\epsilon_{\rm B}$ 
describing the fraction of the internal energy of the internal shock stored in the accelerated electrons 
and the magnetic field respectively. (5) the index, $p_{\rm jet}$,  describing the  
power-law distribution of the electrons in the jet after the acceleration by the shock.
One can refer to the detailed description for the emission of the internal shock  in the  
Appendix of \citet[][]{Yuan2005}.

In the coupled ADAF-jet model, the accretion flow ADAF and the jet are connected by a defined parameter, 
$\eta\equiv\dot M_{\rm jet}/\dot M$, and $\dot M_{\rm jet}$ is input by assuming a value of, $\eta$,
which is free parameter in the present model.
The half-opening angle $\phi$  of the jet in the low/hard state of NS-LMXBs 
is uncertain. In this paper, we fix $\phi=0.1$ as assumed by several other authors for modeling
the SED of the BH-LMXBs \citep[e.g.][]{Yuan2005,Zhang2010}.
Observationally, the bulk Lorentz factor of the jet in the low/hard state of X-ray binaries 
can be restricted in a relatively narrow range and the velocity of the jet is mildly
relativistic, i.e., $\Gamma_{\rm jet}\lesssim 2$. More strictly, the bulk Lorentz factor is
restricted to be as  $\Gamma_{\rm jet}\lesssim 1.67$ \citep[][]{Gallo2003}, 
and $\Gamma_{\rm jet}\lesssim 1.2$ \citep[][]{Fender2006}.
In the internal shock model, the energy density of the internal shock increases with  
increasing $\Gamma_{\rm jet}$, which finally will result in an increase of both the radio emission
and the X-ray emission \citep[][]{Yuan2005}. However, since $\Gamma_{\rm jet}$ is restricted in a very narrow range 
by observations, we expect that a slight change of $\Gamma_{\rm jet}$ will result in a slight  
change of the jet emission.  In the present paper, we fix $\Gamma_{\rm jet}=1.2$ corresponding the 
bulk velocity of the jet $\sim 0.55c$ \citep[][]{Fender2006}. 
The value of $\epsilon_{\rm e}$ and  $\epsilon_{\rm B}$ 
describing the fraction of the internal energy of the internal shock stored in the accelerated electrons 
and the magnetic field, and the index, $p_{\rm jet}$, describing the power-law distribution of the electrons 
in the jet after the acceleration by the shock are uncertain.
\citet[][]{Qiao2015} tested the effect of $\epsilon_{\rm e}$ and $\epsilon_{\rm B}$ 
on the emergent spectrum of the jet in an observationally inferred range of $0.01<\epsilon_{\rm e}<0.1$ and 
$0.01<\epsilon_{\rm B}<0.1$. 
It was found that a change of $\epsilon_{\rm B}$ in the 
range of $\sim 0.01-0.1$, the emergent spectrum of the jet nearly does not change 
(see the right panel of Fig. 3 of \citet[][]{Qiao2015}). A change of  
$\epsilon_{\rm e}$ in the range of $0.01-0.1$, the radio spectrum nearly does not change.
However, the X-ray luminosity changes obviously by changing the value of 
$\epsilon_{\rm e}$ from  $0.01-0.1$ (see the left panel of Fig. 3 of \citet[][]{Qiao2015}). 
As shown in the left panel of Fig. 2 of \citet[][]{Qiao2015}, the X-ray emission 
is completely dominated by the accretion flow (corona) in the luminous X-ray state, which
is also true in the present paper, i.e., the X-ray emission from the ADAF completely dominates the 
X-ray emission from the jet. 
In the present paper, we fix $\epsilon_{\rm e}=0.04$ and 
$\epsilon_{\rm B}=0.02$ respectively throughout the paper as \citep[][]{Qiao2015}.  
The value of the power-law index $p_{\rm jet}$ of the electron 
distribution in the jet predicted by the shock acceleration is $2<p_{\rm jet}<3$.
By modeling the SEDs of three BH-LMXBs, the value of the power-law index $p_{\rm jet}$ of the 
electron distribution is constrained to be 2.1 \citep[][]{Zhang2010}.  Meanwhile, a change of 
$p_{\rm jet}$ in the range of  $2<p_{\rm jet}<3$ has very minor effect on the X-ray spectrum.
In the present paper, we fix the power-law index $p_{\rm jet}=2.1$ throughout the paper.

Finally, we can calculate the emergent spectrum of the coupled ADAF-jet model around a weakly magnetized NS 
by specifying the NS $m$, the NS radius $R_{*}$, the mass accretion rate $\dot m$ and $\eta$.

\section{The results}\label{sec:results}
\subsection{4U 1728-34}\label{sec:4U 1728-34}
We collect the simultaneous radio data and X-ray data of 4U 1728-34, Aql X-1 
and EXO 1743-248 from literatures.  For 4U 1728-34, we get the data from 
\citet[][]{Migliari2003} and \citet[][]{Migliari2006},  in which there are 
12 times simultaneous radio at 8.5 GHz (VLA) and X-ray of 2-10 keV (RXTE) observations.
In the present paper, we only select the hard state observations, so there are 7 observations left.
For some sources observed with type I X-ray burst, especially the sources that showing the
evidence of the photospheric radius expansion (PRE) 
can be used as a distance indicator \citep[][]{vanParadijs1978}. 
The distance of 4U 1728-34 is estimated to be as $d=4.4-4.8$ kpc by taking the minimum peak flux of 
the radius expansion burst as the Eddington limits (assuming $R_{*}=10$ km). 
The uncertainty of the distance arises from the probable range of the NS mass 
$m=1.4-2.0$ \citep[][]{Galloway2003}. 
We convert the flux to the luminosity of 4U 1728-34 by taking the distance 
as $d=4.4$ kpc. One can see the blue open square in the panel (1) of
Fig. \ref{f:sp} for the data.  
The best-fitting linear regression for the correlation between $L_{\rm R}$ and 
$L_{\rm X}$ gives,
\begin{eqnarray}\label{e1:4U 1728-34}
{\rm log} \ L_{\rm R}  =-22.93 + 1.42\times { {\rm log} \ L_{\rm X} }, 
\end{eqnarray}
which can be re-expressed as, $L_{\rm R}=10^{-22.93}L_{\rm X}^{1.42}$.
One can also refer to the blue solid line in the panel (1) of Fig. \ref{f:sp} for clarity. 

In the present paper, we fix the mass and the radius of the NS in 4U 1728-34 as 
$m=1.4$ and $R_{*}=10$ km respectively.
We calculate the emergent spectrum of the 
coupled ADAF-jet model for different $\dot m$, adjusting the value 
of $\eta$ for $L_{\rm X}$ and $L_{\rm R}$ so that 
equation \ref{e1:4U 1728-34} can be satisfied. 
One can see the theoretical points, i.e., the bigger-sized blue solid square in 
the panel (1) of Fig. \ref{f:sp} for details. The corresponding emergent spectra 
can be seen in the panel (2) of Fig. \ref{f:sp}.
Specifically, in the panel (2) of Fig. \ref{f:sp} from the top down, 
the mass accretion rates are $\dot m=3.5\times 10^{-2}$, $\dot m=2.0\times 10^{-2}$, 
$\dot m=1.5\times 10^{-2}$ and $\dot m=8.0\times 10^{-3}$ respectively, and the corresponding $\eta$ are
$\eta(\%)=4.94$, $\eta(\%)=5.03$, $\eta(\%)=5.40$ and $\eta(\%)=3.99$ respectively.
Then a numerical formula of the best-fitting linear regression of $\eta$ as a function of $\dot m$ 
is derived. The formula is as follows,
\begin{eqnarray}\label{e2:4U 1728-34}
{\rm log}\ \eta(\%)=0.92+0.14\times {\rm log}\ \dot m, 
\end{eqnarray}
which can be re-expressed as, $\eta(\%)=8.34\dot m^{0.14}$. One can also refer to the blue solid line 
in Fig. \ref{f:eta} for clarity.

\subsection{Aql X-1}\label{sec:Aql X-1}
We collect the simultaneous radio data and X-ray data of Aql X-1 from \citep[]{Migliari2006,Tudose2009,  
Miller-Jones2010}. We also only  select the observations in the hard state.  
The distance of Aql X-1 is estimated to be as $d=5.2\pm 0.7$ kpc by assuming the 
average peak luminosity of the PRE burst as $\sim 3.8\times 10^{38} \ \rm erg\ s^{-1}$, or 
$\sim 2.0\times 10^{38} \ \rm erg\ s^{-1}$ \citep[][]{Kuulkers2003,Jonker2004}.
In the present paper, we convert the flux to the luminosity by assuming the distance 
of Aql X-1 to be as $d=5.2$ kpc.
One can see the dark green open square in the panel (1) of
Fig. \ref{f:sp} for the data. 
The best-fitting linear regression for the correlation between $L_{\rm R}$ and 
$L_{\rm X}$ gives,
\begin{eqnarray}\label{e1:Aql X-1}
{\rm log} \ L_{\rm R}  =14.8 + 0.38\times { {\rm log} \ L_{\rm X} },
\end{eqnarray}
which can be re-expressed as, $L_{\rm R}=10^{14.8}L_{\rm X}^{0.38}$.
One can also refer to the dark green solid line in the panel (1) of Fig. \ref{f:sp} for clarity. 
In the present paper, we fix the mass and the radius of the NS in Aql X-1 as 
$m=2$\footnote{As discussed in Section \ref{ADAF}, the critical luminosity  
of the ADAF is $L_{\rm crit}\sim 0.1\alpha^2 L_{\rm Edd}=1.26\times 10^{37}m \rm \ erg \ s^{-1}$
(assuming $\alpha=1$). Actually, for Aql X-1 and EXO 1745-248, for some points,
the X-ray luminosity $L_{\rm 2-10 keV }$ has exceeded $10^{37}\rm \ erg \ s^{-1}$.
In order to match the points with higher X-ray luminosities, 
we choose a bigger value of the NS mass, i.e., $m=2$, which is often suggested to be 
the upper limit of the mass of the NS, to fit the radio/X-ray correlation of Aql X-1 and EXO 1745-248.} 
and $R_{*}=10$ km respectively. As in Section \ref {sec:4U 1728-34},
we calculate the emergent spectrum of the coupled ADAF-jet model for 
different $\dot m$, adjusting the value of $\eta$ 
for $L_{\rm X}$ and  $L_{\rm R}$ so that equation \ref{e1:Aql X-1} can be satisfied.  
The theoretical points can be seen as the bigger-sized dark green solid square in 
the panel (1) of Fig. \ref{f:sp}. The corresponding emergent spectra can be seen in the panel (3) of Fig. \ref{f:sp}.
Specifically, in the panel (3) of Fig. \ref{f:sp} from the top down, 
the mass accretion rates are $\dot m=2.5\times 10^{-2}$, $\dot m=1.5\times 10^{-2}$, 
$\dot m=1.0\times 10^{-2}$ and $\dot m=4.0\times 10^{-3}$ respectively, and the corresponding $\eta$ are
$\eta(\%)=3.25$, $\eta(\%)=4.72$, $\eta(\%)=6.25$ and $\eta(\%)=11.6$ respectively.
The best-fitting linear regression of $\eta$ as a function of $\dot m$ is derived. 
The formula is as follows,
\begin{eqnarray}\label{e2:Aql X-1}
{\rm log}\ \eta(\%)=-0.59-0.69\times {\rm log}\ \dot m, 
\end{eqnarray}
which can be re-expressed as, $\eta(\%)=0.26\dot m^{-0.69}$. One can also refer to the 
dark green solid line in Fig. \ref{f:eta} for clarity.

\subsection{EXO 1745-248}\label{sec:EXO 1745-248}
\begin{figure*}
\includegraphics[width=85mm,height=60mm,angle=0.0]{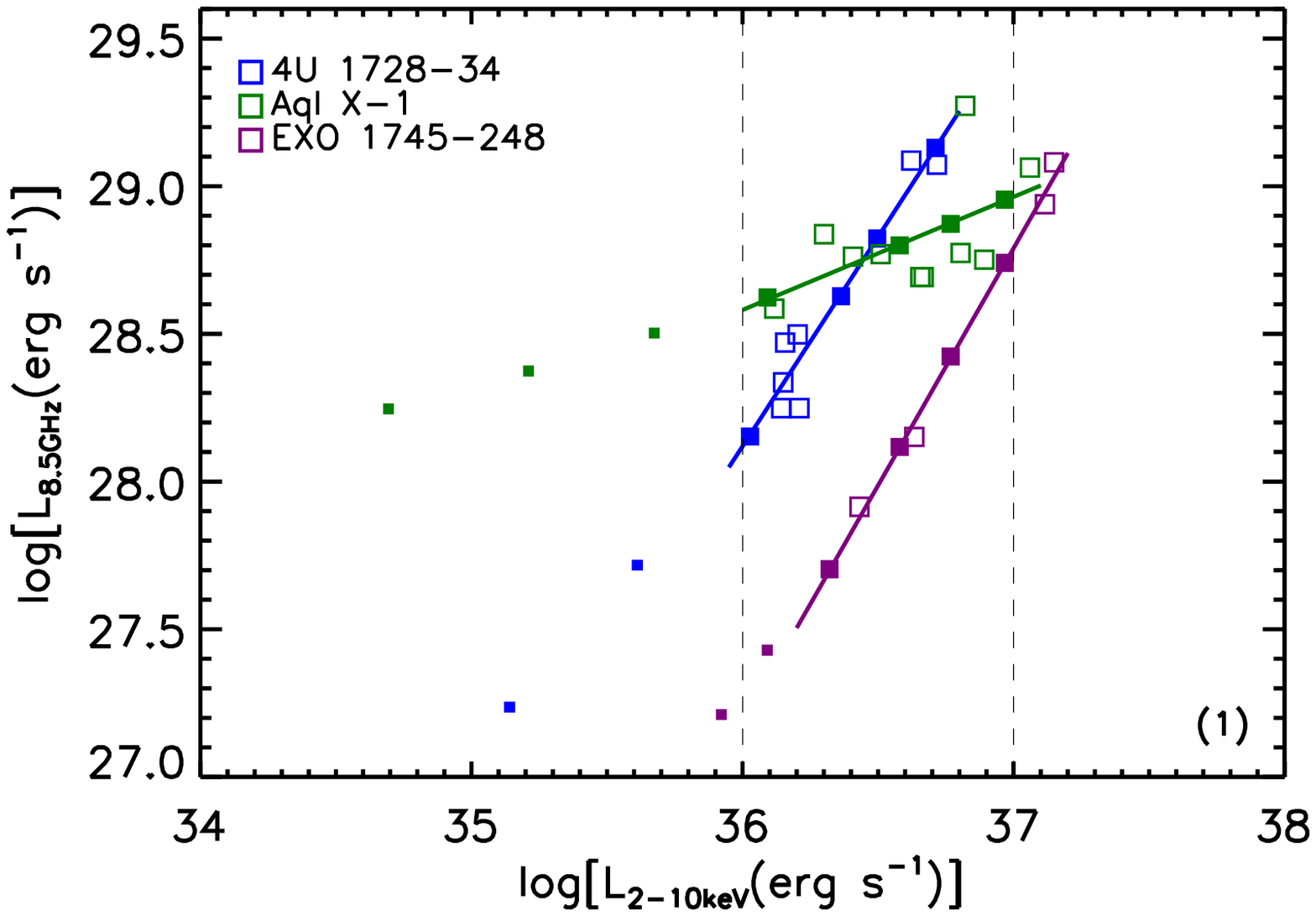}
\includegraphics[width=85mm,height=60mm,angle=0.0]{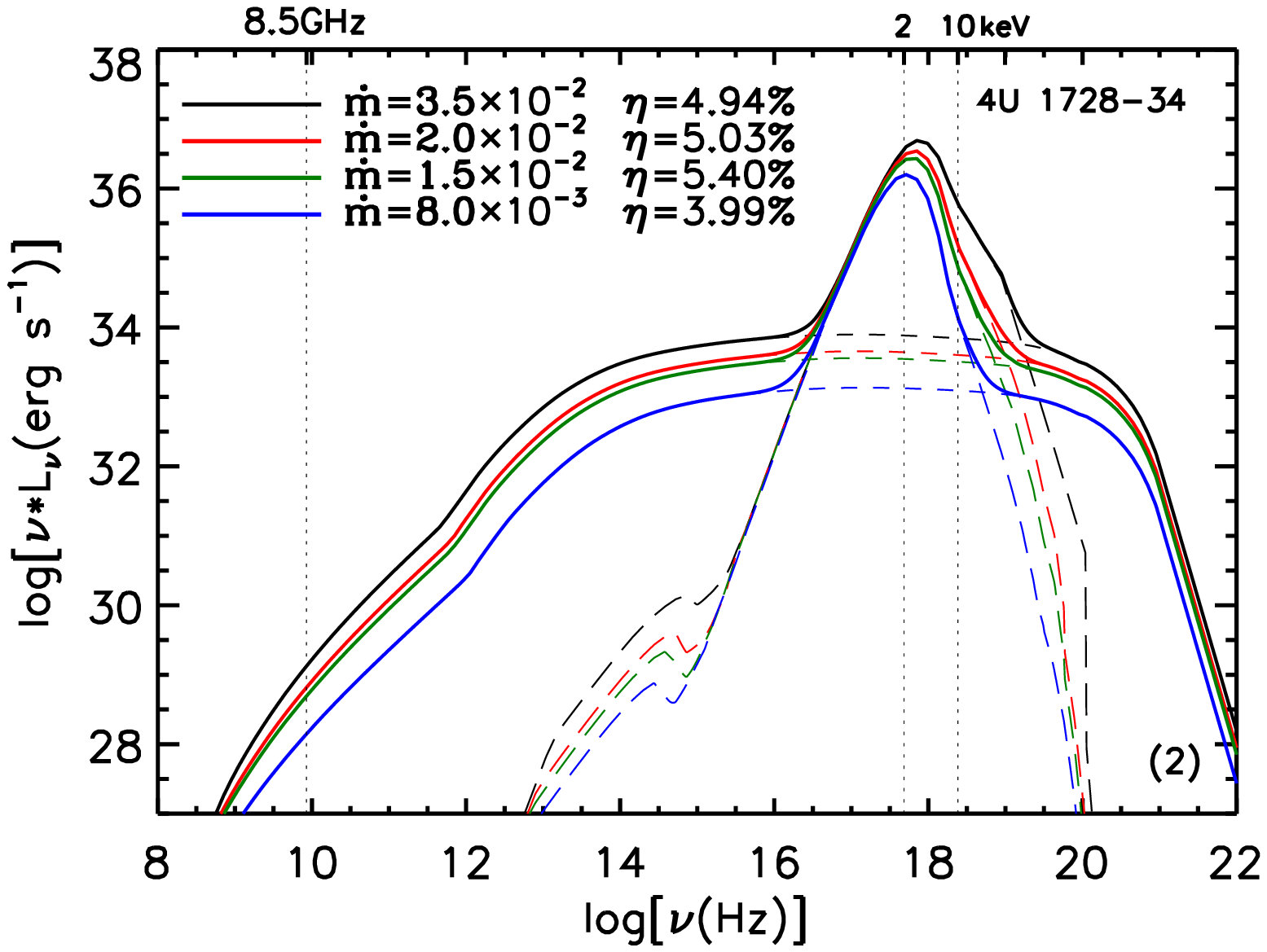}
\includegraphics[width=85mm,height=60mm,angle=0.0]{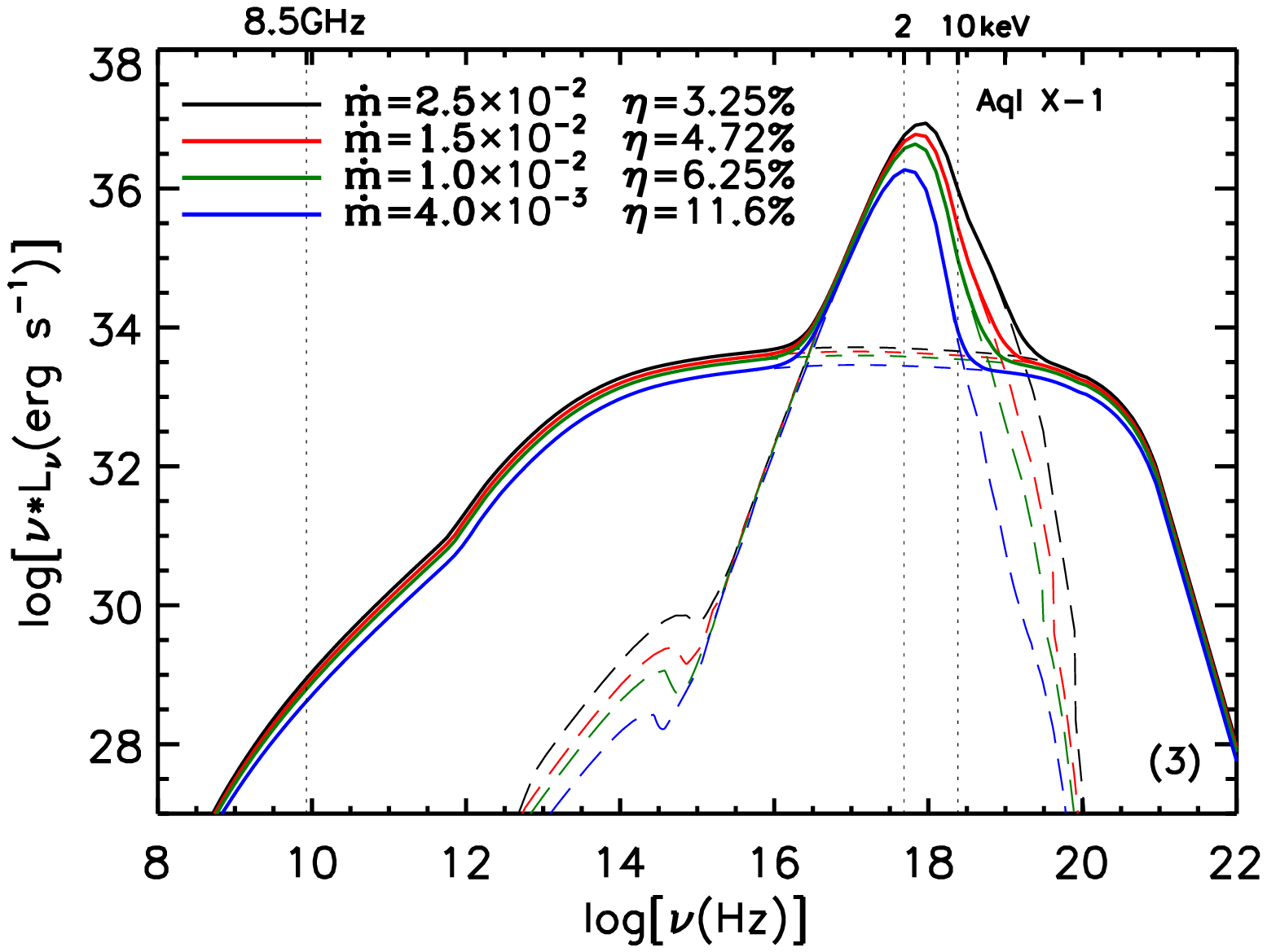}
\includegraphics[width=85mm,height=60mm,angle=0.0]{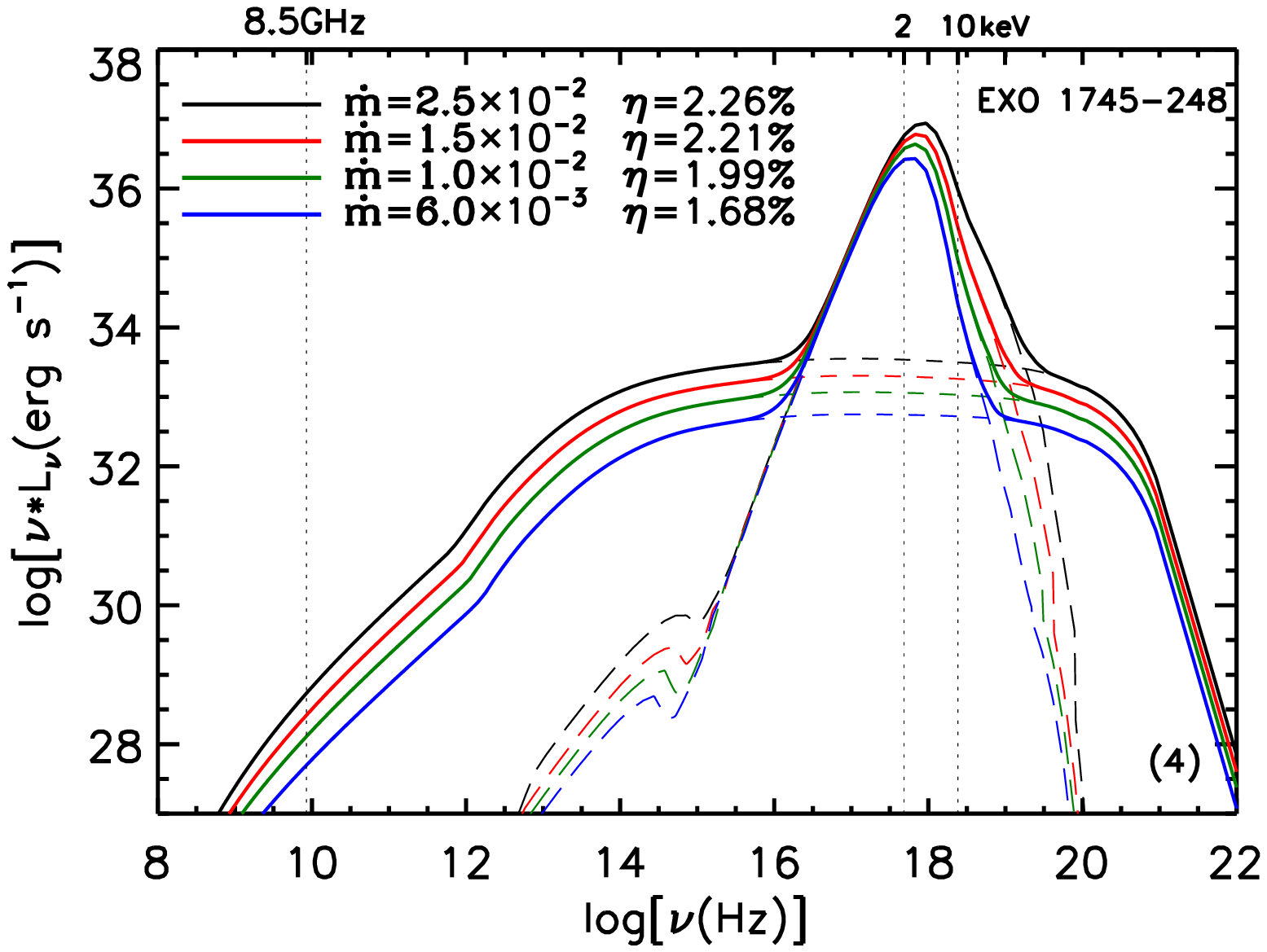}
\caption{\label{f:sp} Panel (1): radio/X-ray correlation of  
4U 1728-34, Aql X-1 and EXO 1745-248. The blue open square, 
the dark green open square and the purple open square
are the observational data of 4U 1728-34, Aql X-1 and EXO 1745-248 respectively.
The blue solid line, the dark green solid line and the purple solid line are
the best-fitting linear regression for the correlation between $L_{\rm R}$ and 
$L_{\rm X}$ of 4U 1728-34, Aql X-1 and EXO 1745-248 respectively.
The bigger-sized blue solid square, the bigger-sized dark green solid square and   
the bigger-sized purple solid square are the theoretical points of the coupled
ADAF-jet model for fitting the radio/X-ray correlation of 
4U 1728-34 (equation \ref{e1:4U 1728-34}), Aql X-1 (equation \ref{e1:Aql X-1}) and 
EXO 1745-248 (equation \ref{e1:EXO 1745-248}) respectively. 
The smaller-sized blue solid square, the smaller-sized dark green solid square and   
the smaller-sized purple solid square are the theoretical points of the coupled
ADAF-jet model for 4U 1728-34 (based on equation \ref{e2:4U 1728-34}), 
Aql X-1 (based on equation \ref{e2:Aql X-1}) and EXO 1745-248 (based on equation \ref{e2:EXO 1745-248}) 
for extrapolating $\eta$ as a function of $\dot m$ down to a lower value of $\dot m$ (see the text for details). 
Panel (2): corresponding emergent spectrum of the ADAF-jet model 
for 4U 1728-34. The solid line is the total emergent spectrum including the ADAF and the jet.
The short-dashed line is the emergent spectrum of the jet,  
and the long-dashed line is  the emergent spectrum of the ADAF.  
Panel (3): corresponding emergent spectrum of the ADAF-jet model 
for Aql X-1. The solid line is the total emergent spectrum including the ADAF and the jet.
The  short-dashed line is the emergent spectrum of the jet,  
and the long-dashed line is the emergent spectrum of the ADAF.  
Panel (4): corresponding emergent spectrum of the ADAF-jet model 
for EXO 1745-248. The solid line is the total emergent spectrum including the ADAF and the jet.
The  short-dashed line is the emergent spectrum of the jet, 
and the long-dashed line is the emergent spectrum of the ADAF.  }
\end{figure*}

EXO 1745-248 located at the globular cluster Terzan 5 is the third NS-LMXB , 
which has been  reasonably  well-established with a correlation between
the radio luminosity and the X-ray luminosity \citep[][]{Tetarenko2016}. 
We collect the near-simultaneous radio data and X-ray data of EXO 1745-248 
in the 2015 outburst \citep[][]{Tetarenko2016}, in which the radio data are from 
VLA, Australia Telescope Compact Array (ATCA), and the X-ray data are from $\it Swift$ 
X-ray telescope.  In the present paper, we convert the flux 
\footnote{In the paper of \citet[][]{Tetarenko2016},  the X-ray luminosity $ L_{\rm X}$ 
in the range of 1-10 keV and the radio luminosity $ L_{\rm R}$ at 10 GHz of EXO 1745-248 are 
used to measure the radio/X-ray correlation. In order to more easily 
to compare with the data of 4U 1728-34 and Aql X-1, we convert the  
X-ray luminosity $ L_{\rm X}$ in the range of 1-10 keV  to 
the range of 2-10 keV  by simply assuming a photon index of $\Gamma=1.5$, and 
convert the radio luminosity $ L_{\rm R}$ at 10 GHz to 8.5 GHz by assuming 
a flat radio spectrum of $f_{\rm \nu}\propto \nu^{\alpha^{'}}$ with $\alpha^{'}=0$. }
to the luminosity by assuming the distance of EXO 1745-248 to be $d=5.9$ kpc 
as that of the distance of the globular cluster Terzan 5 \citep[][]{Valenti2007}.
One can see the purple open square in the panel (1) of
Fig. \ref{f:sp} for the data. 
The best-fitting linear regression for the correlation between $L_{\rm R}$ and 
$L_{\rm X}$ gives,
\begin{eqnarray}\label{e1:EXO 1745-248}
{\rm log} \ L_{\rm R}  =-30.68 + 1.61\times { {\rm log} \ L_{\rm X} },
\end{eqnarray}
which can be re-expressed as, $L_{\rm R}=10^{-30.68}L_{\rm X}^{1.61}$.
One can also refer to the purple solid line in the panel (1) of Fig. \ref{f:sp} for clarity. 
As the case of Aql X-1, in the present paper, we fix the mass and the radius of the NS in 
EXO 1745-248 as $m=2$ and $R_{*}=10$ km respectively. Then we calculate the emergent spectrum of the 
coupled ADAF-jet model for different $\dot m$, adjusting the value 
of $\eta$ for  $L_{\rm X}$ and  $L_{\rm R}$ so that equation \ref{e1:EXO 1745-248} can be satisfied. 
One can see the bigger-sized purple solid square in the panel (1) of Fig. \ref{f:sp}
for the theoretical points. The corresponding emergent spectra can be seen in the panel (4) of Fig. \ref{f:sp}. 
Specifically, in the panel (4) of Fig. \ref{f:sp} from the top down, 
the mass accretion rates are $\dot m=2.5\times 10^{-2}$, $\dot m=1.5\times 10^{-2}$, 
$\dot m=1.0\times 10^{-2}$ and $\dot m=6.0\times 10^{-3}$ respectively, and the corresponding $\eta$ are
$\eta(\%)=2.26$, $\eta(\%)=2.21$, $\eta(\%)=1.99$ and $\eta(\%)=1.68$ respectively.
The best-fitting linear regression of $\eta$ as a function of $\dot m$ is derived. 
The formula is as follows,
\begin{eqnarray}\label{e2:EXO 1745-248}
{\rm log}\ \eta(\%)=0.71+0.21\times {\rm log}\ \dot m, 
\end{eqnarray}
which can be re-expressed as, $\eta(\%)=5.13\dot m^{0.21}$. 
One can also refer to the purple solid line in Fig. \ref{f:eta} for clarity.

As we can see from the panel (2), (3), and (4) of Fig. \ref{f:sp},
it is clear that the radio emission is dominated by the jet, while the X-ray emission is dominated by the
ADAF (energy advected on the surface of the NS) of all the three sources  
4U 1728-34, Aql X-1 and  EXO 1745-248. 

\section{Discussions}
\subsection{The dependence of $\eta$ on $\dot m$: a constraint to jet power}\label{sec:eta}
The relativistic jet is a very common phenomenon associated with BHs or NSs.  
Observationally, there are two kinds of jets in BH X-ray binaries,  i.e., the 
compact jet and the ballistic jet. The compact jet is often  associated with the low/hard state, and the 
ballistic jet is often  associated with the spectral state transition  
between the low/hard state and the high/soft state \citep[][]{Fender2004}.
The compact jet is stable, and less powerful than the ballistic jet. 
So far, several models have been proposed for the formation mechanism of the jet around a BH, e.g.,
the BZ effect by \citet[][]{Blandford1977} and the BP effect by \citet[][]{Blandford1982}.
In the BZ effect, the jet is driven by extracting the rotational energy of the BH
via a large-scale magnetic field. 
While in the BP effect, a large-scale magnetic field threads the accretion disc, extracting 
the rotational energy of the accretion disc to drive the jet. 
\citet[][]{Narayan2012} claimed that there is a correlation between the power of the 
ballistic jet $P_{\rm jet}$ and the spin of the BH $a_{*}$ with a form of  
$P_{\rm jet}\propto a_{*}^2$ (with $a_{*}=cJ/GM^2$, 
$J$ being the angular momentum of the BH, $G$ being the gravitational constant
and $M$ being the mass of the BH) by compiling a sample composed of five  BH X-ray binaries 
with $P_{\rm jet}$ and $a_{*}$ measurements (with $P_{\rm jet}$ estimated by the observed maximum 
radio flux at 5 GHz during the state transition, and $a_{*}$ measured by the continuum-fitting method). 
If the claimed correlation between $P_{\rm jet}$ and $a_{*}$ is true, which supports that 
the ballistic jet may be driven by the BZ process.
However, we should also keep in mind that the claimed correlation between 
$P_{\rm jet}$ and $a_{*}$ is still debatable, which strongly depends on how the 
jet power is estimated and the spin is measured. For example, 
\citet[][]{Russell2013}  systematically studied the dependence of the power of the 
ballistic jets on the spin of 12 BH X-ray binaries, the authors thought that the data 
do not yet support this correlation, which is still needed to be tested by the observations in the future. 
The origin of the compact jet in the low/hard state of BH X-ray binaries is uncertain, 
which is probably driven by the BP process \citep[][for discussions]{Fender2004}.
In both BZ effect and BP effect, the power of the jet is related with
the strength of the large-scale magnetic field \citep[][]{Livio1999}.

\begin{figure}
\includegraphics[width=85mm,height=60mm,angle=0.0]{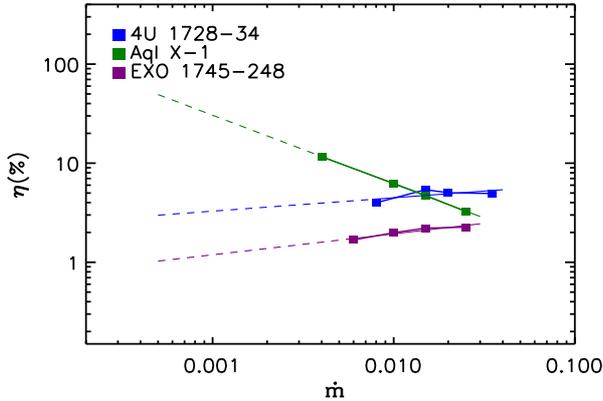}
\caption{\label{f:eta}$\eta$ as a function of $\dot m$.
The blue solid square, the dark green solid square and  the purple solid square  
are the theoretical points derived from the coupled ADAF-jet model  
for fitting the radio/X-ray correlation of  
4U 1728-34 (equation \ref{e1:4U 1728-34}), Aql X-1 (equation \ref{e1:Aql X-1}) and 
EXO 1745-248 (equation \ref{e1:EXO 1745-248}) respectively.
The blue solid line, the dark green solid line and the purple solid line 
are the best-fitting linear regression between $\eta$ and $\dot m$
of the  theoretical points of 4U 1728-34, Aql X-1 and EXO 1745-248 respectively.
The blue dashed line, the dark green dashed line and the purple dashed line
are the extrapolating part of $\eta$ as a function of $\dot m$ down to a lower value of 
$\dot m$ for 4U 1728-34 (based on equation \ref{e2:4U 1728-34}), Aql X-1 (based on equation \ref{e2:Aql X-1})  
and EXO 1745-248 (based on equation \ref{e2:EXO 1745-248}) respectively.  }
\end{figure}

Observationally, in the low/hard state, the power of the jet around a NS is much lower than
that of a BH. Specifically, for a fixed  X-ray luminosity, the radio luminosity of a
NS is generally 1-2 orders of magnitude lower than that of a BH \citep[e.g.][]{Corbel2013,Tudor2017,Gallo2018}.
The study of $\eta$ as function $\dot m$ in the present paper may provide some clues  
for the jet power in the low/hard state of NS.
As can be seen in Section \ref{sec:4U 1728-34}, Section \ref{sec:Aql X-1},
Section \ref{sec:EXO 1745-248}, we calculate the emergent spectrum of the 
coupled ADAF-jet model at different $\dot m$, adjusting the value of $\eta$ 
for $L_{\rm X}$ and $L_{\rm R}$ to fit the radio/X-ray correlation of 4U 1728-34, Aql X-1 
and  EXO 1745-248 respectively. Then, we derive a fitting formula between $\eta$ and $\dot m$,
i.e., $\eta(\%)=8.34\dot m^{0.14}$, $\eta(\%)=0.26\dot m^{-0.69}$, and 
$\eta(\%)=5.13\dot m^{0.21}$ for 4U 1728-34, Aql X-1, and EXO 1745-248 respectively.
One can refer to the blue solid line for 4U 1728-34, the dark green solid line for Aql X-1, 
and the purple solid line for EXO 1745-248 in Fig. \ref{f:eta} for clarity. 
If the derived fitting formula between $\eta$ and $\dot m$ can be extrapolated
down to a lower value of $\dot m$, we can predict a new value of X-ray luminosity 
and radio luminosity. One can see the blue dashed line, the dark green dashed line 
and the purple dashed line in Fig. \ref{f:eta} for the extrapolating part of $\eta$ as a function of  
$\dot m$ for 4U 1728-34, Aql X-1 and EXO 1745-248 respectively.
For 4U 1728-34, we take $\dot m=4\times 10^{-3}$ and $\dot m=2\times 10^{-3}$ as examples.
The numerical results of $\eta(\%)$, and the corresponding predicted X-ray luminosity $L_{\rm X}$
and the radio luminosity $L_{\rm R}$ are shown in Table \ref{predict}.
One can also refer to the smaller-sized blue solid square in the panel (1) 
of Fig. \ref{f:sp} for clarity.
For Aql X-1, we take $\dot m=2\times 10^{-3}$,
$\dot m=1\times 10^{-3}$ and $\dot m=5\times 10^{-4}$ as examples.
The numerical results of $\eta(\%)$, and the corresponding $L_{\rm X}$
and $L_{\rm R}$ are shown in Table \ref{predict}.
One can also refer to the smaller-sized dark green solid square in the panel (1) 
of Fig. \ref{f:sp} for clarity.
For EXO 1745-248, we take $\dot m=4\times 10^{-3}$ and 
$\dot m=3\times 10^{-3}$ as examples.
The numerical results of $\eta(\%)$, and the corresponding $L_{\rm X}$
and $L_{\rm R}$ are shown in Table \ref{predict}.
One can also refer to the smaller-sized purple solid square in the panel (1) 
of Fig. \ref{f:sp} for clarity.
We expect that such the predicted X-ray luminosity and the corresponding radio luminosity 
for 4U 1728-34, Aql X-1 and EXO 1745-248 could be confirmed by the observations in the future.  

\begin{table}
\caption{Extrapolating value of $\eta$ (based on equation \ref{e2:4U 1728-34} for 4U 1728-34, equation \ref{e2:Aql X-1} 
for Aql X-1 and equation \ref{e2:EXO 1745-248} for EXO 1745-248), and the corresponding  
X-ray luminosity $L_{\rm X}$ and the radio luminosity $L_{\rm R}$ 
based on the ADAF-jet model for 4U 1728-34, Aql X-1 and EXO 1745-248 respectively.}
\centering
\begin{tabular}{ccccccccc}
\hline\hline
4U 1728-34 \\
\hline
$\dot m$ &  $\eta(\%)$  & $L_{\rm X} (\rm \ erg \ s^{-1})$ & $L_{\rm R} (\rm \ erg \ s^{-1})$   \\
\hline
$4\times 10^{-3}$  & 3.85   & $4.1\times 10^{35}$  & $5.2\times 10^{27} $    \\
$2\times 10^{-3}$  & 3.49   & $1.4\times 10^{35}$  & $1.7\times 10^{27} $    \\
\hline\hline
Aql X-1\\
\hline
$\dot m$ &  $\eta(\%)$  & $L_{\rm X} (\rm \ erg \ s^{-1})$ & $L_{\rm R} (\rm \ erg \ s^{-1})$   \\
\hline
$2\times 10^{-3}$  & 18.9   & $4.7\times 10^{35}$  & $3.2\times 10^{28} $    \\
$1\times 10^{-3}$  & 30.5   & $1.6\times 10^{35}$  & $2.4\times 10^{28} $    \\
$5\times 10^{-4}$  & 49.3   & $4.9\times 10^{34}$  & $1.8\times 10^{28} $    \\
\hline\hline
EXO 1745-248 \\
\hline
$\dot m$ &  $\eta(\%)$  & $L_{\rm X} (\rm \ erg \ s^{-1})$ & $L_{\rm R} (\rm \ erg \ s^{-1})$   \\
\hline
$4\times 10^{-3}$  & 1.61   & $1.2\times 10^{36}$  & $2.7\times 10^{27} $    \\
$3\times 10^{-3}$  & 1.51   & $8.4\times 10^{35}$  & $1.6\times 10^{27} $    \\
\hline\hline
\end{tabular}
\\
\label{predict}
\end{table}

The formation mechanism of the jet around a NS is uncertain.
One of the possible mechanisms for the jet formation is the 
`propeller' effect \citep[][]{Illarionov1975}.
In the `propeller' model, when the matter in the accretion disc is accreted towards the NS, 
the matter is halted at the NS magnetosphere in the case of magnetic field 
rotating locally at super-Keplerian speed. The magnetic pressure of the 
magnetosphere will balance the ram pressure of the accreted matter, accelerating and  
ejecting a fraction of the matter in the inner region of the disc around 
the magnetosphere forming the jet. 
The `propeller' effect  works at the lower luminosity regime, and has been used to explain the
emissions in the quiescent state of several Be/X-ray pulsars \citep[][]{Tsygankov2017}. 
As an example, `propeller' jet is supported by the detection of the radio emission 
after the fast X-ray decay during its outburst in 1998 in NS SAX J1808.4-3658 
\citep[][]{Gaensler1999}. The jet power predicted by the `propeller' model is also 
uncertain, depending on the NS spin, magnetic field, accretion rate etc. \citep[e.g.][for discussions]{Tudor2017}. 
We should keep in mind that the theoretical explanation for the relatively fainter 
radio emission in NSs compared with that of in BHs is still a question to be answered. 
The studies in the present paper show that $\eta$ decreases slightly with decreasing $\dot m$ for 
4U 1728-34 and EXO 1745-248, while $\eta$ increases with decreasing $\dot m$ for Aql X-1.
One can see Fig. \ref{f:eta} for clarity.
It is also clear that although $\eta$ as a function of $\dot m$  has similar trends between
4U 1728-34 and EXO 1745-248, the value of $\eta$ for 4U 1728-34 is roughly three times higher than that 
of EXO 1745-248. We note that in a wide range of $\dot m$, the value of $\eta$
for Aql X-1 is greater than that of 4U 1728-34 and EXO 1745-248.
It is very possible that the origin of the such discrepancy of 
$\eta$ as a function of $\dot m$ in 4U 1728-34, EXO 1745-248 and Aql X-1 is resulted by
the different strength of the large-scale magnetic field among them. 
Actually, a coherent millisecond X-ray pulsation (at $\sim 10^{35} \rm \ erg \ s^{-1}$) 
has been discovered in Aql X-1, while the X-ray pulsation was not discovered in 4U 1728-34 and EXO 1745-248, 
supporting that Aql X-1 may have a relatively stronger 
large-scale magnetic field compared with that of 4U 1728-34 and EXO 1745-248
for channeling the matter onto the magnetic poles \citep[][]{Casella2008}.
 
\subsection{The narrow X-ray luminosity range of the radio/X-ray correlation}\label{narrow}
As we can see from Section \ref{sec:results}, for NS-LMXBs, so far, the correlation between the radio 
luminosity and the X-ray luminosity has been  reasonably well-established only in three sources 4U 1728-34, 
Aql X-1 and  EXO 1745-248.  Meanwhile, we should keep in mind that 
such a radio/X-ray correlation holds only in a very narrow X-ray luminosity range ($\sim 1$ dex)
between $\sim 10^{36} \rm \ erg \ s^{-1}$ and $\sim 10^{37} \rm \ erg \ s^{-1}$.
As discussed in Section \ref{sec:eta}, currently it is unclear whether the established radio/X-ray 
correlation of 4U 1728-34, Aql X-1 and  EXO 1745-248 in the hard state can be extended down to 
the X-ray luminosity regime $\lesssim 10^{36} \ \rm erg\ s^{-1}$.
We expect that our predicted X-ray luminosity and the corresponding radio luminosity could 
be confirmed by the future observations.
Actually, whether NS can still launch jet at lower mass accretion rates 
(corresponding $L_{\rm X} \lesssim 10^{36} \ \rm erg\ s^{-1}$) is uncertain. 
\citet[][]{Tudor2017} investigated the radio property and the X-ray property of 
three accreting millisecond X-ray pulsars (AMXPs), 
IGR J17511-3057, SAX J1808.4-3658 and IGR J00291+5934 during their outbursts in 2015,
as well as a non-pulsing NS Cen X-4 in the  low-luminosity state 
($L_{\rm X} \lesssim 10^{36} \ \rm erg\ s^{-1}$) of 2015 and the outburst in 1979. 
The authors found that only IGR J00291+5934 and SAX J1808.4-3658 have radio detections
when they are in the low-luminosity state. 
Further, the authors showed that there is a complicated correlation between the radio luminosity
and the X-ray luminosity in SAX J1808.4-3658 and there is a tight correlation 
between the radio luminosity and the X-ray luminosity in IGR J00291+5934.
Recently, \citet[][]{Gallo2018} collected a sample composed of 41 NSs with 
hard state Atolls and AMXPs, including three transitional millisecond pulsars (tMSPs)
with the X-ray luminosity down to $\sim 10^{31}\ \rm erg\ s^{-1}$.
Collectively, it is found that there is a correlation between the radio luminosity and 
the X-ray luminosity with the slope $\beta=0.44^{+0.05}_{-0.04}$.
While separately the slope is  $\beta=0.71^{+0.11}_{-0.09}$ for hard state Atolls, 
$\beta=0.27^{+0.09}_{-0.10}$ for AMXPs,  
$\beta=1.16^{+0.28}_{-0.24}$ for AMXPs weighted Atolls, and  
$\beta=1.39^{+0.35}_{-0.30}$ for AMXPs-tMSPs weighted Atolls respectively. 
In the present paper, we focus only on the individual NS-LMXBs for their 
radio/X-ray correlation respectively. 
Meanwhile, NS-LMXBs generally have a relatively lower magnetic field ($\lesssim 10^{8}\ \rm G$),
which is consistent with the weakly magnetized ADAF model we used in the present paper. 
The interpretation of the radio/X-ray correlation in 
AMXPs and tMSPs for $L_{\rm X} \lesssim 10^{36} \ \rm erg\ s^{-1}$
will be continued in the future work within the framework of the modified 
ADAF (with stronger large-scale magnetic field considered)-jet model, 
which exceeds the scope of the present paper.   

\subsection{The effect of the NS spin}
In the present paper, we consider that the internal energy stored in the ADAF and the radial 
kinetic energy of the ADAF are transferred onto the surface of the NS as in \citet[][]{Qiao2018b}. 
Meanwhile, it is assumed that all this energy, i.e., $f_{\rm th}=1$, is thermalized at the surface of 
the NS, and then radiated out as the blackbody emission to be scattered in the ADAF itself. 
Actually, more generally, the energy transfered onto the surface of the NS should include not only
the internal energy and radial kinetic energy of the ADAF but also the rotational energy of the ADAF.
The rotational energy of the ADAF will be released in a very thin boundary layer between the accretion 
flow and the surface of the NS. In the Newtonian approximation, the rotational energy released in the 
boundary layer can be expressed as,
\begin{eqnarray}\label{e:BL}
L_{\rm BL}={1\over 2}{\dot M} {GM\over R_{*}} k \bigg(1-{\nu_{\rm NS}\over \nu_{*}}\bigg)^2,
\end{eqnarray}
where $L_{\rm BL}$ is the rotational energy released in the boundary layer, 
$M$ is the mass of the NS, $R_{*}$ is the radius of the NS, 
$k$ is a correction factor in the range of 0-1 ($k=1$ for the thin disc),
$\nu_{\rm NS}$ is the rotational frequency (spin) of the NS, 
and $\nu_{*}$ is the rotational frequency of the ADAF at $R_{*}$. 
From equation \ref{e:BL}, it is clear that $L_{\rm BL}$ is related with the spin frequency 
of the NS $\nu_{\rm NS}$. If we take a typical value of $k={1\over3}$, for $\nu_{\rm NS}=0$, 
$L_{\rm BL}={1\over 6}{\dot M} {GM\over R_{*}}$, meaning $1\over 6$ of the gravitational energy
will be released in the boundary layer. For $\nu_{\rm NS}=\nu_{*}$, 
the rotational energy released in the boundary layer is $L_{\rm BL}=0$. 
The radiation from the boundary layer will also be scattered in the ADAF, changing the structure
and the X-ray emission of the ADAF. \citet[][]{Burke2018} compiled a sample composed of nine 
NSs (not accreting pulsars) with well determined spin frequency. By fitting the X-ray spectra, 
the authors found that indeed the spin frequency of the NS is correlated with some X-ray related quantities, 
such as the electron temperature in corona, the Compton $y$-parameter, and the Compton amplification factor etc. 
Meanwhile, as discussed in Section \ref{narrow} for the `propeller' effect, theoretically, 
the spin of NS plays a very important role in the formation and the power of the jet. 
The observational hint for the correlation
between the spin frequency and the jet power in NSs was confirmed by \citet[][]{Migliari2011}.
A more systematic study of the effects of the spin of NS on 
the X-ray emission from the ADAF and the radio emission from the jet (jet power) for the 
radio/X-ray correlation in NSs will be continued in the future work.

\section{Conclusions}
In this paper, we explain the radio/X-ray correlation of three NS-LMXBs, i.e.,  
$L_{\rm R}\propto L_{\rm X}^{\sim 1.4}$ for 4U 1728-34,
$L_{\rm R}\propto L_{\rm X}^{\sim 0.4}$ for Aql X-1, and 
$L_{\rm R}\propto L_{\rm X}^{\sim 1.6}$ for EXO 1745-248 in their hard state 
within the framework of the coupled ADAF-jet model respectively. 
By modelling the observed radio/X-ray correlation, we derive a fitting formula of $\eta$ as a 
function of $\dot m$ for 4U 1728-34, Aql X-1 and EXO 1745-248 respectively.
We extrapolate the dependence of $\eta$ on $\dot m$ to a lower value of $\dot m$, it is found that
in a wide range of $\dot m$, the value of $\eta$ in Aql X-1 is greater than that of   
in 4U 1728-34 and EXO 1745-248. The relatively higher value of $\eta$ in Aql X-1 
compared with that of in 4U 1728-34 and EXO 1745-248 implies that Aql X-1 may have a  
relatively higher magnetic field, which is supported by the discovery of the coherent millisecond
X-ray pulsation (at $\sim 10^{35} \rm \ erg \ s^{-1}$) in Aql X-1.
Finally, we predict a new X-ray luminosity and radio luminosity based on the relation between $\eta$
and $\dot m$ extrapolating down to a lower value of $\dot m$.  
We expect that the predicted X-ray luminosity and radio luminosity
can be confirmed by the future observations, with which we may give some constraints 
to the formation mechanism and the power of the jet for 
4U 1728-34, Aql X-1, and EXO 1745-248 respectively.

\section*{Acknowledgments}
This work is supported by the National Natural Science Foundation of
China (Grants 11773037 and 11673026), the gravitational wave pilot B (Grants No. XDB23040100), 
and the National Program on Key Research and Development Project (Grant No. 2016YFA0400804).

\bibliographystyle{mnras}
\bibliography{qiaoel}


\bsp	
\label{lastpage}
\end{document}